\documentclass[12pt]{article}

\usepackage[dvips]{graphicx}

\newcommand{\nn}{\nonumber}
\newcommand{\be}{\begin{equation}}
\newcommand{\ee}{\end{equation}}
\newcommand{\ba}{\begin{eqnarray}}
\newcommand{\ea}{\end{eqnarray}}
\newcommand{\req}[1]{(\ref{#1})}
\def\={\,=\,}
\newcommand{\ci}[1]{\cite{#1}}

\def\vk{{\bf k}_{\perp}}
\def\vkk{k^2_\perp}

\def\vbs{{\bf b}}
\def\vb0{{\bf b}_0}

\def\als{\alpha_s}

\def\mev{\,{\rm MeV}}
\def\gev{\,{\rm GeV}}

\def\xbj{x_{\rm Bj}}

\newcommand{\wf}{wave function}
\newcommand{\lsim}{\raisebox{-4pt}{$\,\stackrel{\textstyle
                                                         <}{\sim}\,$}}
\newcommand{\gsim}{\raisebox{-4pt}{$\,\stackrel{\textstyle
                                                         >}{\sim}\,$}}

\newcommand{\tw}{\textwidth}
\def\xb{\bar{x}}

\def\veps{\varepsilon}

\begin{document}
\thispagestyle{empty}
\begin{flushright}
WU B 08-05 \\
September, 23 2008\\[20mm]
\end{flushright}

\begin{center}
{\Large\bf The target asymmetry in hard vector-meson electroproduction
  and parton angular momenta} \\
\vskip 15mm

S.V.\ Goloskokov
\footnote{Email:  goloskkv@theor.jinr.ru}
\\[1em]
{\small {\it Bogoliubov Laboratory of Theoretical Physics, Joint Institute
for Nuclear Research,\\ Dubna 141980, Moscow region, Russia}}\\
\vskip 5mm

P.\ Kroll \footnote{Email:  kroll@physik.uni-wuppertal.de}
\\[1em]
{\small {\it Fachbereich Physik, Universit\"at Wuppertal, D-42097 Wuppertal,
Germany}}\\
and\\
{\small {\it Institut f\"ur Theoretische Physik, Universit\"at
    Regensburg, \\D-93040 Regensburg, Germany}}\\

\end{center}

\vskip5mm
\begin{abstract}
\noindent The target asymmetry for electroproduction of vector mesons
is investigated within the handbag approach. While the generalized
parton distribution (GPD) $H$ is taken from a previous analysis of the
elctroproduction cross section, we here construct the GPD $E$ 
from double distributions and constrain it by the Pauli form factors of
the nucleon, positivity bounds and sum rules. Predictions for the
target asymmetry are given for various vector mesons and discussed how
experimental data on the asymmetry  will further constrain $E$ and
what we may learn about the angular momenta the partons carry.   
\end{abstract}

\section{Introduction}
\label{sec:intro}
The handbag approach to hard exclusive reactions which bases on
factorization into hard subprocesses and soft GPDs, attracted the
interest of many theoreticians and 
experimentalists in the last decade. The still poorly known GPDs as
well as the quality and scantiness of the experimental data prevented 
definite conclusions about the applicability of the handbag approach
in the experimentally accessible region of kinematics so far. 
Now, the situation is changing; we are in a stage where an increasing 
amount of precise data on hard exclusive reactions becomes available. 
The already accumulated data from JLAB, HERMES, COMPASS and HERA as well
as those to be expected in the near future, will allow for an 
extraction of a wealth of qualitative and quantitative information on
the GPDs. In a few years from now we will likely have accumulated
sufficient information on the GPDs in order to answer the question in
which range of kinematics the handbag approach can be applied to hard
exclusive reactions in a consistent way. 

Here in this work, we are going to investigate the target asymmetry
$A_{UT}$ for electroproduction of flavor neutral vector mesons
($\rho^0$, $\omega$, $\phi$) and for the the processes
$ep\to e\rho^+n$ and $ep\to e K^{*0}\Sigma^+$. The asymmetry is
related to the imaginary part of an interference term between the two
GPDs $H$ and $E$. Provided $H$ is sufficiently well-known from an
analysis of, say, the unpolarized cross section for electroproduction 
of vector mesons \ci{GK1,GK2}, one may extract information on $E$ from 
data on $A_{UT}$. Admittedly this is only possible in a model-dependent 
way, i.e.\ one unavoidably has to exploit an ansatz for $E$ with a few 
free parameters which can be adjusted to experiment. Although the data 
on the target asymmetry will likely suffer from large experimental 
errors (cf.\ the recent, still preliminary HERMES result for $\rho^0$
production \ci{ami})  we believe and are going to substantiate this hope
in the following, that the pattern of future $A_{UT}$ data for various vector 
mesons will likely render such a determination of $E$ feasible. 
Provided this program can be successfully carried through, one may evaluate 
Ji's sum rule \ci{ji:97} from $H$ and $E$ and learn about the angular
momenta the partons inside the proton possess. The possibility of
extracting information on the quark angular momenta from $A_{UT}$, at
least for $u$ and $d$ quarks has been discussed by Ellinghaus {\it et al} 
\ci{nowak} first. Here in this work, we will investigate improved
parameterizations of the GPD $E$ as compared to \ci{nowak} and will also
study the role of $E$ for gluons and strange quarks.

In the next section we will briefly sketch the handbag formalism for
the processes of interest. In Sect.\ 3 we will present the model we use
for the GPD $E$ and discuss the constraints on its parameters. 
Predictions for $A_{UT}$ are presented and discussed in Sect.\ 4.
Concluding remarks will be presented in Sect.\ 5.

\section{The handbag formalism}
\label{sec:handbag}    
For the analysis of the target asymmetry for vector-meson ($V$)
electroproduction we restrict ourselves to the kinematical region of
small skewness ($\xi\lsim 0.1$) and small invariant momentum transfer 
($-t\lsim 0.5\,\gev^2$) but large photon virtuality 
($Q^2\gsim 3\,\gev^2$) and large energy in the photon-proton center of 
mass frame ($W\gsim 5\,\gev$). In this kinematical region we have
already investigated the $\gamma^* p\to Vp$ cross sections for 
unpolarized protons \ci{GK1,GK2,GK3} and flavor neutral vector mesons 
and achieved very good agreement with the available data from HERMES, 
Fermilab and HERA. These cross sections are strongly dominated by 
contributions from the GPD $H$; it is safe to neglect~\footnote{ 
We remark that in the unpolarized cross section there are no
interference terms between $H$ and the other GPDs up to corrections of
order $\xi^2$.}
the other GPDs $E$, $\widetilde{H}$ and $\widetilde{E}$. In our
previous work we calculated the quark ($\gamma^* q\to Vq$) and gluon
($\gamma^* g\to Vg$) subprocess amplitudes within the modified
perturbative approach \ci{botts89} in which quark transverse
degrees of freedom as well as Sudakov suppressions are taken into
account in the subprocess. This approach allows us to calculate not
only the asymptotically dominant (longitudinal) amplitude for 
$\gamma^*_Lp\to V_Lp$ but also the one for transversely polarized
photons and vector mesons ($\gamma^*_Tp\to V_Tp$). In contrast to the 
longitudinal amplitude the latter one cannot be calculated in
collinear approximation since it suffers from infrared singularities
in this limit \ci{man,teryaev}. The quark transverse momenta, $\vk$, 
provide an admittedly model-dependent regularization scheme of these 
singularities by replacements of the  type 
\be
    \frac1{d Q^2} \longrightarrow \frac1{d Q^2 + \vkk}
\ee
in the parton propagators. Here, $d$ is a momentum fraction or a
product of two.       
  
The dominant helicity amplitudes for the process $\gamma^* p\to VB$ are given by
\ba
{\cal M}_{\mu +, \mu +}(V) &=& \frac{e}{2}\left\{ \sum_a e_a {\cal C}_V^{aa}
\langle H\rangle^g_{V\mu} + \sum_{ab} {\cal C}_V^{ab} 
\langle H\rangle^{ab}_{V\mu}\right\}\,,\nn\\
{\cal M}_{\mu -, \mu +}(V) &=& -\frac{e}{2}\frac{\sqrt{-t}}{M+m}\left\{ 
\sum_a e_a {\cal C}_V^{aa} \langle E\rangle^g_{V\mu} + \sum_{ab} {\cal C}_V^{ab} 
\langle E\rangle ^{ab}_{V\mu}\right\}\,,
\label{hel-amp}
\ea
where some simplifications, relevant for the small $\xi$ region, have
been used. Thus, for instance, the contribution from $E$ to the
helicity non-flip amplitude, being proportional to $\xi^2$, is neglected.
The quark flavors are denoted by $a, b$ while $e_{a(b)}$
denotes the quark charges in units of the positron charge $e$ and $m$
the proton mass. Besides the proton we consider also other ground
state baryons, $B$, as for instance the $\Sigma^+$ (with mass $M$). 
With the help of flavor symmetry the $p\to B$ transition GPDs can be
related to the proton ones \ci{Frankfurt:99}.
The weight factors ${\cal C}_V^{ab}$ comprise the flavor structure of
the mesons. The non-zero weight factors for selected vector
mesons read
\ba
{\cal C}_{\rho^0}^{\,uu}&=&-{\cal C}_{\rho^0}^{\,dd} \= {\cal C}_\omega^{\,uu} 
\={\cal C}_\omega^{\,dd}\=1/\sqrt{2}\,, \qquad {\cal C}_\phi^{\,ss}\=
1\,, \nn\\
{\cal C}_{\rho^+}^{\,ud} &=& 1\,, \qquad {\cal C}_{K^{*0}}^{\,ds}\=1\,,
\label{flavor}
\ea

The explicit helicities in \req{hel-amp} refer to the proton while $\mu$
is the helicity of the photon and the meson. Only the $t$ dependence of
the GPDs is taken into account in the amplitudes \req{hel-amp}. That of the
subprocess amplitudes ${\cal H}$ provides corrections of order $t/Q^2$
which we neglect throughout this paper. In contrast to the subprocess
amplitudes the $t$ dependence of the GPDs is scaled by a soft parameter, 
actually by the slope of the diffraction peak. It can be shown
\ci{GK1} that in addition to the familiar parity-invariance relation 
the following symmetry relation 
\be
{\cal M}_{-\mu\nu^\prime,-\mu\nu} \= 
             {\cal M}_{\mu\nu^\prime,\mu\nu}
\label{sym-N}
\ee
holds among the helicity amplitudes \req{hel-amp}. Helicity flips in
the $\gamma^*\to V$ transition have to be generated within the
subprocess. Hence, amplitudes involving such helicity flips are
suppressed by powers of $\sqrt{-t}/Q$ and consequently neglected. 

The terms $\langle F\rangle$ denote convolutions of subprocess
amplitudes and a GPD $F (=H,E)$. For the gluonic subprocess the
convolutions read
\be
\langle F \rangle_{V\mu}^g \= \sum_\lambda\int_0^1 d\xb 
   {\cal H}^{Vg}_{\mu\lambda,\mu\lambda}(\xb,\xi,Q^2,t=0) F^g(\xb,\xi,t)\,,
\ee  
where the label $\lambda$ refers to the helicities of the partons
participating in the subprocess. The subprocess amplitudes ${\cal H}$
are discussed in great detail in Ref.\ \ci{GK1}. We refrain from
repeating the lengthy expressions here. For the quark subprocesses we have 
\be
\langle F \rangle_{V\mu}^{ab} \= \sum_\lambda\int_{-1}^1 d\xb  
  {\cal H}^{Vab}_{\mu\lambda,\mu\lambda}(\xb,\xi,Q^2,t=0)\,F^{ab}(\xb,\xi,t)\,.   
\ee
The quark GPDs for the proton and the $p\to n$ and $p\to\Sigma^+$ transitions
are
\be
F^{aa}\= F^a\,, \qquad F^{ab}\=F^a-F^b \quad (a\neq b)\,.
\ee
 
Since the resummation of the logarithms involved in the Sudakov factor
can only be performed in the impact parameter space efficiently
\ci{botts89} we quote the subprocess amplitudes in that space
\ba
{\cal H}^{Vab}_{\mu\lambda,\mu\lambda} &=& \int d\tau d^2b\, 
         \hat{\Psi}_{V\mu}(\tau,-\vbs)\, 
      \hat{\cal F}^{ab}_{\mu\lambda,\mu\lambda}(\xb,\xi,\tau,Q^2,\vbs)\, \nn\\ 
      && \times   \als(\mu_R)\,{\rm exp}{[-S(\tau,\vbs,Q^2)]}\,.
\label{mod-amp}
\ea
For the Sudakov factor $S$, the choice of the renormalization
($\mu_R$) and factorization ($\mu_F$) scales as well as the hard
scattering kernels $\hat{\cal F}$ or their respective Fourier
transforms ${\cal F}$, we refer to Ref.\ \ci{GK3}. In order to
generalize to the case of flavored mesons the following replacements
in quark propagators occuring in ${\cal F}$ have to be made
\be
T_s \longrightarrow e_a T_s\,, \qquad T_u \longrightarrow e_b T_u\,.
\ee
The last item in \req{mod-amp} to be explained is $\hat{\Psi}_{V\mu}$,
the Fourier transform of the momentum-space light-cone \wf{} for the
vector meson. It is parameterized as a simple Gaussian ($j=L,T$) 
\ba 
\Psi_{Vj}(\tau,\vk) &=& 8\pi^2\sqrt{2N_c} f_{Vj}(\mu_F) a_{Vj}^2 
                 \Big[1+B_1^{Vj}(\mu_F) C_1^{3/2}(2\tau-1)  \nn\\
           &+& B_2^{Vj}(\mu_F) C_2^{3/2}(2\tau-1)\Big]\,
                   {\rm exp}{[-a_{Vj}^2\vk^2/(\tau\bar{\tau})]}\,.
\label{wavefunction}
\ea
The parameters of the $\rho$ and $\phi$ \wf s are specified in
\ci{GK3}. The decay constants of longitudinally polarized $\omega$ and
$K^{*0}$ mesons are 187 and $218\,\mev$, respectively. For the ratio
of $f_{VT}$ and $f_{VL}$ we take the QCD sum rule value of $\simeq 0.8$ 
at the scale of $1\,\gev$ \ci{braun96}. The Gegenbauer coefficients 
$B_1^{Vj}$ are zero for all vector mesons except for the $K^{*0}$ for 
which the values 0 and 0.1 are chosen. For the second Gegenbauer 
coefficients we take $B_2^{VL}=0$ and $B_2^{VT}=0.1$. All values of
the Gegenbauer coefficients which are quoted at the scale $1\,\gev$, 
are in agreement with recent QCD sum rule analyses \ci{braun96,ball07},
only $B_2^{VL}$ is slightly smaller even with regard of the errors of
the QCD sum rule results quoted in \ci{braun96}. The evolution of
the decay constants for transversely polarized vector mesons 
as well as that of the Gegenbauer coefficients with the 
factorization scale \ci{shifman} is taken into account. Finally, for 
the transverse size parameters we take the values 
$a_{\omega j}=a_{K^{*0} j}=a_{\rho^+ j}=a_{\rho^0 j}$. 
The  values of the latter parameters as well as those for the $\phi$
meson can be found in \ci{GK3}. We note in passing that Gaussian 
\wf s of the type \req{wavefunction} are frequently used in phenomenology.

There is a minimal value of $-t$ allowed in the process of interest
\be
t_{\rm min}\= -\frac{2\xi}{1-\xi^2}\,\big[(1+\xi)M^2-(1-\xi)m^2\big]\,.
\ee
Given the smallness of the $\Sigma^+$-$p$ mass difference $t_{\rm
  min}$ is small in all cases, practically of order $\xi^2$ and is
therefore neglected as other effects of this order. We note  
that our helicities are light-cone ones which naturally occur in the 
handbag approach. The differences to the usual c.m.s. helicities are
of order $m\sqrt{-t}/W^2$ \ci{diehl03} and can be ignored in the 
kinematical region of interest in this work. Skewness is related to 
Bjorken-$x$, $\xbj$, by
\be
\xi\simeq \frac{\xbj}{2-\xbj}\Big[1+m_V^2/Q^2\Big]\,,
\label{xi-xbj}
\ee
where $m_V$ denotes the mass of the vector meson. 
The helicity amplitudes are normalized such that the partial cross
sections for $\gamma^*_{L(T)}p\to V_{L(T)}p$ read ($\Lambda$ is the 
usual Mandelstam function)
\be
\frac{d\sigma_{L(T)}}{dt} \= \frac1{16\pi (W^2-m^2) 
         \sqrt{\Lambda(W^2,-Q^2,m^2)}}\,\sum_{\nu^\prime}
                           |{\cal M}_{0(+)\nu^\prime,0(+)+}|^2\,,
\label{sigma}
\ee
which holds with regard to the above-mentioned simplifications. For
most of the processes of interest the proton helicity flip amplitude
can be neglected in \req{sigma}; the only exception is the $\rho^+$
channel. The cross sections, integrated upon $t$, are denoted by 
$\sigma_L$ and $\sigma_T$. The full (un-separated) cross section for 
$\gamma^*p\to Vp$ is
\be
\sigma \= \sigma_T + \veps \sigma_L\,,
\ee
in which $\varepsilon$ is the familiar ratio of longitudinal to transverse
photon fluxes. The power corrections of kinematical origin included in 
Eq.\ \req{xi-xbj} and in the phase space factor \req{sigma} are taken 
into account by us. With the exception of these kinematical effects 
hadron masses are omitted otherwise.

Following the conventions specified in \ci{sapeta} (see also
\ci{kugler07}) the target asymmetry for $\gamma^* p\to
VB$ reads
\be
A_{UT} \= -2\; \frac{{\rm Im}\big[{\cal M}^{*}_{+-,++}\,{\cal
      M}_{++,++}\big]
       +\veps\, {\rm Im}\big[{\cal M}^{*}_{0-,0+}\,{\cal
         M}_{0+,0+}\big]}
        {\sum_{\nu^\prime}\Big[\mid {\cal M}_{+\nu^\prime,++}\mid^2 
        + \veps \mid {\cal M}_{0\nu^\prime,0+}\mid^2\Big]}\,.
\label{eq:AUT}
\ee 
It is is measured as the $\sin(\phi-\phi_S)$ moment of the cross section 
for electroproduction of vector meson with a transversally polarized
proton target where $\phi$ is the azimuthal angle between the lepton
and hadron plane and $\phi_S$ the azimuthal angle of the target spin 
vector defined with respect to the lepton plane \ci{sapeta}. 

\section{Modeling $E$}
\label{sec:E}
\subsection{Valence quarks}
\label{subsec:val}
Not much is known about $E$ as yet. Some information on
it comes from the zero-skewness analysis \ci{DFJK4} of the Pauli form
factor of the nucleon. This electromagnetic form factor is odd under 
charge conjugation and therefore only sensitive to valence quarks 
($q_{\rm val}=q-\bar{q}$). The forward limit $t=\xi=0$ of $E_{\rm val}$ 
is parameterized like that of $H$, i.e.\ the familiar parton distribution
functions (PDFs),
\ba
e^{a}_{\rm val}(x) &=& E^{a}_{\rm val}(x,\xi=0,t=0) \nn\\ 
   &=& \frac{\Gamma(2-\alpha_{\rm val}+\beta^a_{\rm val})}{\Gamma(1-\alpha_{\rm val})
\Gamma(1+\beta^a_{\rm val})}\,\kappa_a\, x^{-\alpha_{\rm
    val}(0)}\,(1-x)^{\beta^{\,a}_{\rm val}}\,.
\label{e-ansatz}
\ea
The moments of $e^a_{\rm val}$ read~\footnote{
Generally the forward limit of a GPD $F$ is defined as 
$f^a(x)=F^a(x,\xi=0,t=0)$ for quarks and $xf^g(x)=F^g(x,\xi=0,t=0)$ 
for gluons ($x\geq 0$). The $n$-th moment of $f^i$ ($i=a,g$) is
defined as $f_{n0}^i=\int_0^1 dx x^{n-1}f^i(x)$. This definition 
holds for antiquarks too.}
\be
e^{a_v}_{n0} \= \int_0^1 dx x^{n-1} e^{a}_{\rm val} (x) \= \kappa_a 
\frac{\Gamma(2-\alpha_{\rm val}+\beta^{\,a}_{\rm
    val})}{\Gamma(1+n-\alpha_{\rm val}+\beta^{\,a}_{\rm val})}
\frac{\Gamma(n-\alpha_{\rm val})}{\Gamma(1-\alpha_{\rm val})}\,.
\label{e-moments}
\ee
As can be seen from this expression the pre-factor in \req{e-ansatz} 
ensures the normalization
\be
e^{a_v}_{10} \= \kappa_a\,,
\ee
where $\kappa_a$ is the contribution of flavor-$a$ quarks to the
anomalous magnetic moment of the proton ($\kappa_u \simeq 1.67$,
$\kappa_d \simeq -2.03$). The forward limit \req{e-ansatz} of $E$ can
be used as input to the familiar double distribution ansatz \ci{mus99}
\be
f^{\,a}_{\rm val}(\beta,\alpha,t)\= {\rm e}^{b^{\,e}_{\rm
    val}t}\,\beta^{-\alpha_{\rm val}'\,t}\, 
         e^{a}_{\rm val}(\beta)\, \frac34\,\frac{[(1-\beta)^2-\alpha^2]}
                           {(1-\beta)^{3}} \Theta(\beta)\,.
\label{DD}
\ee 
The GPD is subsequently obtained from the integral
\be
E^{a}_{\rm val}(\xb,\xi,t)\=\int_{-1}^1 d\beta\,\int_{-1+|\beta|}^{1-|\beta|} d\alpha\,
                 \delta(\beta+\xi\alpha-\xb)\,f^{\,a}_{\rm val}(\beta,\alpha,t)\,.
\label{GPD-DD}
\ee
The $t$ dependence in \req{DD} is a small-$t$ simplification of the
more complicated profile function used in
\ci{DFJK4}. $\alpha_{\rm val}(t)=\alpha_{\rm val}(0) + \alpha_{\rm val}'t$ 
is a standard Regge trajectory for which we take the numerical values 
$\alpha_{\rm val}(0)=0.48$ and $\alpha'_{\rm val}=0.9\,\gev^{-2}$ and 
$b_{\rm val}^{\,e}$ is a parameter that describes the $t$ dependence of
the Regge residue. In accord with \ci{GK3} we take $b^{\,e}_{\rm val}=0$. 
Possible dependencies of the Regge trajectory and $b^{\,e}_{\rm val}$ on 
the quark flavor is ignored by us. In the form factor analysis 
performed in \ci{DFJK4} for instance a value of $0.38\,\gev^{-2}$ for 
$b^{\,e}_{\rm val}$ has been found for $u$-valence and
$-0.75\,\gev^{-2}$ for $d$-valence quarks. Our value may be viewed as a rough
average of these values.

In Ref.\ \ci{DFJK4} the powers $\beta^{\,a}_{\rm val}$ have been
determined. The best fit values which hold at a scale of $Q_0=2\gev$, are
\be
\beta^{\,u}_{\rm val}\= 4\,, \qquad \beta^{\,d}_{\rm val} \= 5.6\,.
\label{default}
\ee
However, $\beta^u_{\rm val}$ can be varied between 4 and 6,
$\beta^d_{\rm val}$ between 5 and 6 and still reasonable fits of the 
Pauli form factors are achieved~\footnote{
The analysis of the Pauli form factor suffers from the large number of
free parameters. In contrast to the analysis of the Dirac form factor 
the forward limit $e^{\,a}_{\rm val}$ has to be determined as well. 
Hence, not all parameters can be freed in the fits to the Pauli form 
factors.}.
These parameter regions are taken into account in our error assessment
to be discussed below. Fits with $\beta^{\,u}_{\rm val}$ substantially 
larger than $\beta^{\,d}_{\rm val}$ lead to results of lesser, just 
tolerable quality.  
 
There is a remarkable feature of the parameterization \req{e-ansatz}. 
With the help of \req{e-moments} one can write the ratio of sum and 
difference of the second moments as
\be
\frac{e^{u_v}_{20}+e^{d_v}_{20}}{e^{u_v}_{20}-e^{u_v}_{20}} \=
\frac{\kappa_u +\kappa_d}{\kappa_u-\kappa_d}\, 
\frac{1+\frac{\kappa_u}{\kappa_u+\kappa_d}\,
\frac{\beta^{\,d}_{\rm val}-\beta^{\,u}_{\rm val}}{2-\alpha_{\rm
    val}+\beta^{\,u}_{\rm val}}}
{1+\frac{\kappa_u}{\kappa_u-\kappa_d}\,
\frac{\beta^{\,d}_{\rm val}-\beta^{\,u}_{\rm val}}{2-\alpha_{\rm
    val}+\beta^{\,u}_{\rm val}}}\,.
\label{ratio-moments}
\ee
This expression tells us that for $\beta^{\,d}_{\rm val}>\beta^{\,u}_{\rm val}$ 
this combination of moments is smaller than the corresponding ratio of
the first moments
\be
\frac{\kappa_u +\kappa_d}{\kappa_u-\kappa_d} \= -0.097\,.
\label{ratio-first}
\ee
On the other hand, for $\beta^{\,u}_{\rm val}>\beta^{\,d}_{\rm val}$ 
\req{ratio-moments} is larger than \req{ratio-first}. Note that in the 
chiral soliton model one finds $e^{u_v}_{20}+e^{d_v}_{20}=0$ in the 
large-$N_c$ limit \ci{ossmann}. In the limit of an infinitely large 
factorization scale QCD predicts $\sum_a e^{\,a}_{20}=0$ as Ji showed 
\ci{ji:95}. These two results support to some extent the findings of 
Ref.\ \ci{DFJK4} that the Pauli form factor favors $\beta^{\,d}_{\rm val}\geq
\beta^{\,u}_{\rm val}$. Finally we remark that $E^a_{\rm val}$ constructed
from $e^{\,a}_{\rm val}$ through the double distribution ansatz \req{DD}, 
\req{GPD-DD}, respects the inequalities which ensure the positivity of
the quark densities for various combinations of proton and quark spins 
\ci{DFJK4,pobyl,burkardt03}.

\subsection{Gluons and sea quarks}

In order to model $E$ for gluons and sea quarks we follow Diehl and
Kugler \ci{kugler07} and have recourse to positivity bounds and to a
sum rule that follows from a combination of Ji's sum rule \ci{ji:97}
and the momentum sum rule known from deep inelastic lepton-nucleon 
scattering
\be
e^g_{20} \= -\sum_a e_{20}^{a_v} - 2 \sum_a e_{20}^{\bar{a}}\,.
\label{E-sum-rule}
\ee 
For obvious reasons we split the quark GPDs into valence and sea
quark contributions (using $E^{\bar{q}}(x,\xi,t)=-E^q(-x,\xi,t)$).
As we discussed in the Sect.\ \ref{subsec:val} the analysis of the
Pauli form factor favors a very small value of the valence quark term
in \req{E-sum-rule}. Thus, the gluon and sea quark moments have to
cancel each other almost completely. Assuming in analogy to the
situation for $H$, small sea quark contributions, we have to conclude 
from these considerations that the gluon moment at $t=0$ is very
small, in fact smaller than the individual valence quark moments. 
Another argument that points into the same direction comes from the 
Regge behaviour of $E^g$. As is well-known the familiar soft Pomeron
exchange dominantly couples to the proton helicity non-flip vertex
while its flip coupling is very small. In fact, it is hard to find
convincing evidence for a non-zero flip coupling phenomenologically 
\ci{donnachie}. Supposing that this behavior also holds for the hard
gluonic Regge exchange controlling $H^g$ and $E^g$ at low-$x$, we are 
forced to infer that $E^g\simeq 0$. Thus, the relative importance of 
gluon and valence quark GPDs are likely very different for $E$ and $H$ 
for which $H^g$ is large. On account of this one may neglect $E$ for
gluons and sea quarks in a first attempt and estimate $A_{UT}$ and spin 
density matrix elements (SDMEs) for a transversely polarized proton 
target just from $E^a_{\rm val}$.

In order to elucidate the role of  $E^g$ and $E^{\rm sea}$ 
in more detail we also study scenarios for which these GPDs are
non-zero. Again we employ the double distribution ansatz for them~\footnote{
Possible $D$ terms \ci{polyakov99} in $E$ are identical to those
occuring in $H$ where we have neglected them \ci{GK1,GK2} and, 
therefore, in $E$ too.}. 
The corresponding forward limits are parameterized in analogy to 
\req{e-ansatz} 
\ba
e^s &=& N_s x^{-1-\delta}\,(1-x)^{\beta^s}\,, \nn\\
e^g &=& N_g x^{-1-\delta}\,(1-x)^{\beta^g}\,, 
\ea
For simplicity a flavor symmetric sea is assumed. The low-$x$ behavior 
of both these GPDs is assumed to be controlled by the gluon
trajectory for which we use $\delta=0.1+0.06\ln{(Q^2/Q_0^2)}$ and
$\alpha'_g=0.15\,\gev^{-2}$ for its slope. The gluon trajectory has
been fixed in our analysis of the cross sections for $\rho^0$ and 
$\phi$ electroproduction \ci{GK2,GK3}. 

Further help in fixing the gluon and sea quark GPD comes from the 
positivity bounds.  For an exponential $t$ dependence of the
corresponding double distributions with profile functions $g_i(x)$ 
for $E^i$ and $f_i(x)$ for $H^i$ \ci{GK2,GK3} (cf.\ \req{DD})
\ba
g_s &=& g_g \= \alpha'_g \ln{1/x} + b^{\,e}_g\,, \nn\\
f_s &=& f_g \= \alpha'_g \ln{1/x} + b_g\,,
\label{profile}
\ea
one finds the following bounds \ci{DFJK4}
\ba
\left[\frac{e_s(x)}{s(x)}\right]^2 &\leq&
                  21.75m^2\left[\frac{g_s(x)}{f_s(x)}\right]^3\,
                                   \Big[f_s(x)-g_s(x)\Big]\,,\nn\\
\left[\frac{e_g(x)}{g(x)}\right]^2 &\leq&
                  21.75m^2\left[\frac{g_g(x)}{f_g(x)}\right]^3\,
                                   \Big[f_g(x)-g_g(x)\Big]\,.
\label{bounds}
\ea
These bounds ensure positive semi-definite densities of partons in the
transverse plane \ci{burkardt03}. Contributions from the polarized
PDFs are neglected in \req{bounds} which appears reasonable for not
too large values of $x$ since there is growing experimental evidence
that $\Delta g$ and $\Delta s$ are very small
in that region of $x$, see for instance \ci{hermes-xx,compass-xx}.                                
Inserting the profile functions \req{profile} into the bounds
\req{bounds}, we find  ($i=s,g$)
\be
\beta^i\geq 6\,, \qquad b^{\,e}_g<b_g\,.
\ee
The present data do not provide any other information on $b^{\,e}_g$. We
therefore follow \ci{kugler07} and assume $b^{\,e}_g=0.9 b_g$. The
parameter $b_g$ is taken from our previous work \ci{GK2,GK3}. It has
the value 
\be
b_g\=2.58\,\gev^{-2} + 0.25\,\gev^{-2} \ln{[m^2/(Q^2+m^2)]}\,.
\ee
The restriction of $\beta^i$ guarantees the dominance of the valence
quarks at large $x$. 

We investigate the following variants of $E$ (see Tab.\ \ref{tab:1}): 
Besides our default case, the variant 1, with the powers \req{default}
and $E^g\simeq E^s \simeq 0$ we try three more examples: For variant
2 (with $N_s>0$) and 3 (with $N_s<0$) we choose $\beta^{\,s}=7$ and $\mid
N^s\mid$ as large as the bound \req{bounds} allows. In fact its
saturation occurs at $x\simeq 0.1$. For these variants we choose 
$\beta^{\,g}=6$ and fix $N_g$ by \req{E-sum-rule}. Variant 4 is an
extreme case for which we assume a large value of $\beta^{\,u}_{\rm val}$ 
and a small value of $\beta^{\,d}_{\rm val}$. For this choice the
Pauli form factor of the proton is still fitted although with a low
quality~\footnote{
The JLab measurement E02-13 of the electric form factor of the neutron for
$Q^2$ up to $3.5\,\gev^2$ will tell us whether or not the choice 
$\beta^u_{\rm val} >\beta^d_{\rm val}$ is realistic. The data are not yet available.}.
The valence quarks now contribute substantially to the sum rule 
\req{E-sum-rule}. We assume $E^s=0$ for this variant, take
$\beta^{\,g}=7$ and fix $N_g$ by \req{E-sum-rule} again.
One may also consider a variant where $E$ for the valence quarks is
the same as for variant 4 but trying to saturate the sum rule
\req{E-sum-rule} solely by the sea quarks. However, this would require a
very large sea-quark GPD which violates the positivity bound
\req{bounds}. Finally we investigate the case $\beta^{\,u}_{\rm
  val}= \beta^{\,d}_{\rm val}=6$ and saturate the sum rule
\req{E-sum-rule} either by the gluon (variant 5) or by the sea quarks
(variant 6). In Tab.\ \ref{tab:1} the parameters of the six variants
are compiled. It is to be emphasized that for all variants we consider the gluonic
GPD $E^g$ is far below the bound \req{bounds} since the gluon PDF
$g(x)$ is very large. 
 
\begin{table*}[t]
\renewcommand{\arraystretch}{1.4} 
\begin{center}
\begin{tabular}{|c|c|c|| c |c | c | c ||c |c|c|c|}
\hline     
 var.   & $\beta^{\,u}_{\rm val}$ & $\beta^{\,d}_{\rm val}$ & $\beta^g$ & $\beta^s$ &
 $N_g$ & $N_s$ & $J^u$ & $J^d$ & $J^s$ & $J^g$ \\[0.2em]   
\hline
1 & 4 &  5.6 & - & - & 0.000  & 0.000  & 0.250 & 0.020  & 0.015  & 0.214 \\[0.2em]
2 & 4 &  5.6 & 6 & 7 & -0.873 & 0.155  & 0.276 & 0.046  & 0.041  & 0.132 \\[0.2em]
3 & 4 &  5.6 & 6 & 7 & 0.776  & -0.155 & 0.225 & -0.005 & -0.011 & 0.286 \\[0.2em]
4 & 10&  5   & 7 & - & 0.523  & 0.000  & 0.209 & 0.013  & 0.015  & 0.257 \\[0.2em]
5 & 6 &  6   & 7 & - & 0.167  & 0.000  & 0.230 & 0.024  & 0.015  & 0.228  \\[0.2em]
6 & 6 &  6   & - & 7 & 0.000  & 0.025  & 0.234 & 0.028  & 0.019  & 0.214 \\[0.2em]
\hline
\end{tabular}
\end{center}
\caption{The parameters of the forward limits of the GPD $E$ and the
  angular momentum the various partons carry. $\alpha=0.48$
  throughout.  
  The parameters and the angular momenta are quoted at a scale of $2\,\gev$.}
\label{tab:1}
\renewcommand{\arraystretch}{1.0}   
\end{table*}

\subsection{Parton angular momenta}
In Tab.\ \ref{tab:1} we also quote the angular momenta of the quarks 
and the gluons evaluated from Ji's sum rule \ci{ji:97} 
\ba
\langle J^a \rangle &=& \frac12\,\Big[ e^{a_v}_{20} +
  h^{a_v}_{20}\Big] + e^{\bar{a}}_{20} + h^{\bar{a}}_{20}\,, \nn\\
\langle J^g \rangle &=& \frac12\, \Big[ e^{g}_{20} + h^{g}_{20}\Big]\,.  
\label{angular-mom}
\ea
The second moments of $H$ at $\xi=t=0$ are evaluated from the CTEQ6
PDFs \ci{cteq6}. The following values have been obtained:
\ba
h^{u_v}_{20} &=& 0.288\,, \quad h^{d_v}_{20}\=0.118\,, \quad h^{g}_{20} \=0.428\,,\nn\\
h^{\bar{u}}_{20} &=& 0.028\,, \quad h^{\bar{d}}_{20} \= 0.035\,, 
\quad h^{s}_{20} \= 0.015\,,
\label{h20-moments}
\ea
at a scale of $2\,\gev$. Note that $\langle J^a\rangle$ is the
average three-component of the angular momentum quarks of flavor $a$
and their antiquarks carry. The angular momenta of the various partons
do not sum exactly to $1/2$, the spin of the proton, since the moments
quoted in \req{h20-moments} do not exactly saturate the momentum sum rule of
deep inelastic lepton-nucleon scattering (actually $\sum_i h^i_{20}=0.990$) 
because of the neglected charm distribution.  
The values quoted in Tab.\ \ref{tab:1} reveal a characteristic pattern:
for all the variants we investigate the angular momenta of the $u$
quarks and the gluon are large while those of the $d$ and $s$ quarks
are small. Hence, the spin of the proton is essentially made up
by the angular momenta of the $u$ quarks and the gluons. In Ref.\
\ci{DFJK4} further solutions for $E^u_{\rm val}$ and $E^d_{\rm val}$
are given which are also compatible with the data on the Pauli form factor
of the proton. These solutions, completed if necessary by $E^g$ and/or
$E^s$ in the same fashion as we constructed the variants quoted in
Tab.\ \ref{tab:1}, provide results for the angular momenta which lie 
between the extreme values of the variants shown in the table.
The present theoretical uncertainties of the angular momenta are indicated 
by the spread of the values for $J^i$ quoted in Tab.\ \ref{tab:1}. 

The angular momenta of the valence quarks are
\be
\langle J^{u_v}\rangle\= 0.222\,, \qquad \langle J^{d_v}\rangle\= -0.015\,,
\ee
for the variants 1,2 and 3. Slightly different values are obtained for
the other variants. These values agree well with those derived
in \ci{DFJK4} within the errors estimated in the latter work. The
largest deviation occurs for variant 4 for which the angular momenta of
the valence quarks are about 1.5 standard deviations smaller than the
results found in \ci{DFJK4} ($\langle J^{u_v}\rangle= 0.211(17)$,
$\langle J^{d_v}\rangle=0.000(19)$). In Ref.\ \ci{haegler} the angular
momenta of $u$ and $d$ quarks have been calculated from lattice QCD using
domain wall valence quarks and improved staggered sea quarks in heavy quark
scenarios. After extrapolation to the physical value of the pion mass
they obtain
\be
\langle J^u \rangle\= 0.214 (27)\,, \qquad \langle
J^d\rangle\=-0.001 (27) \,.
\ee
It should be stressed that these results are rather to be interpreted
as the angular momenta of the valence quarks although after the
extrapolation they may contain sea quark effects. Obviously, our
values for the valence quarks agree with the lattice results. The
angular momenta of the gluons have been estimated using QCD sum rules 
\ci{balitsky97} and a quark model \ci{barone98}. The values found 
($\langle J^g\rangle \simeq 0.25$ \ci{balitsky97} and $\simeq 0.24$ 
\ci{barone98} at the respective low scales of 1 and $0.5\,\gev$, lie 
within the range of results quoted in Tab.\ \ref{tab:1} even though on 
the side of larger values.
                                            
The orbital angular momentum of the partons may be obtained from 
\req{angular-mom} by subtracting the corresponding first moments of 
$\widetilde{H}$. In Ref.\ \ci{BB} (scenario 1, NLO) the moments of the
polarized valence quarks PDFs are evaluated to  
\be
\tilde{h}^{u_v}_{10} \=\phantom{-} 0.926\,, \qquad
\tilde{h}^{d_v}_{10}\=-0.341\,, 
\label{axial-vector-BB}
\ee 
at the scale of $2\,\gev$. 
Just for orientation we quote the values of the orbital angular
momenta of the valence quarks for variants 1,2 and 3 at our default scale
\be
\langle L^{u_v}\rangle \simeq -0.241\,, \quad \langle L^{d_v}\rangle \simeq 0.155\,.
\ee
In the analysis of Ref.\ \ci{BB} the first moment of the polarized
gluon distribution, $x\Delta g(x)$, is subject to huge errors and
cannot be used here. Because of the smallness of $x\Delta g(x)$ at 
least for small $x$ \ci{hermes-xx,compass-xx} we expect 
$\tilde{h}^g_{10}\simeq 0$ and, hence, $\langle L^g\rangle \simeq \langle J^g\rangle$.
We note in passing that in a recent analysis of the polarized PDFs
\ci{florian} a gluon moment  $\tilde{h}^g_{10}$ has been found that is
compatible with zero at least at the scale of $3.16\,\gev$.

\section{Discussion of the results on $A_{UT}$}
\label{sec:results}

\begin{figure}[ht]
\includegraphics[width=0.45\tw,bb=36 346 508 743,clip=true]{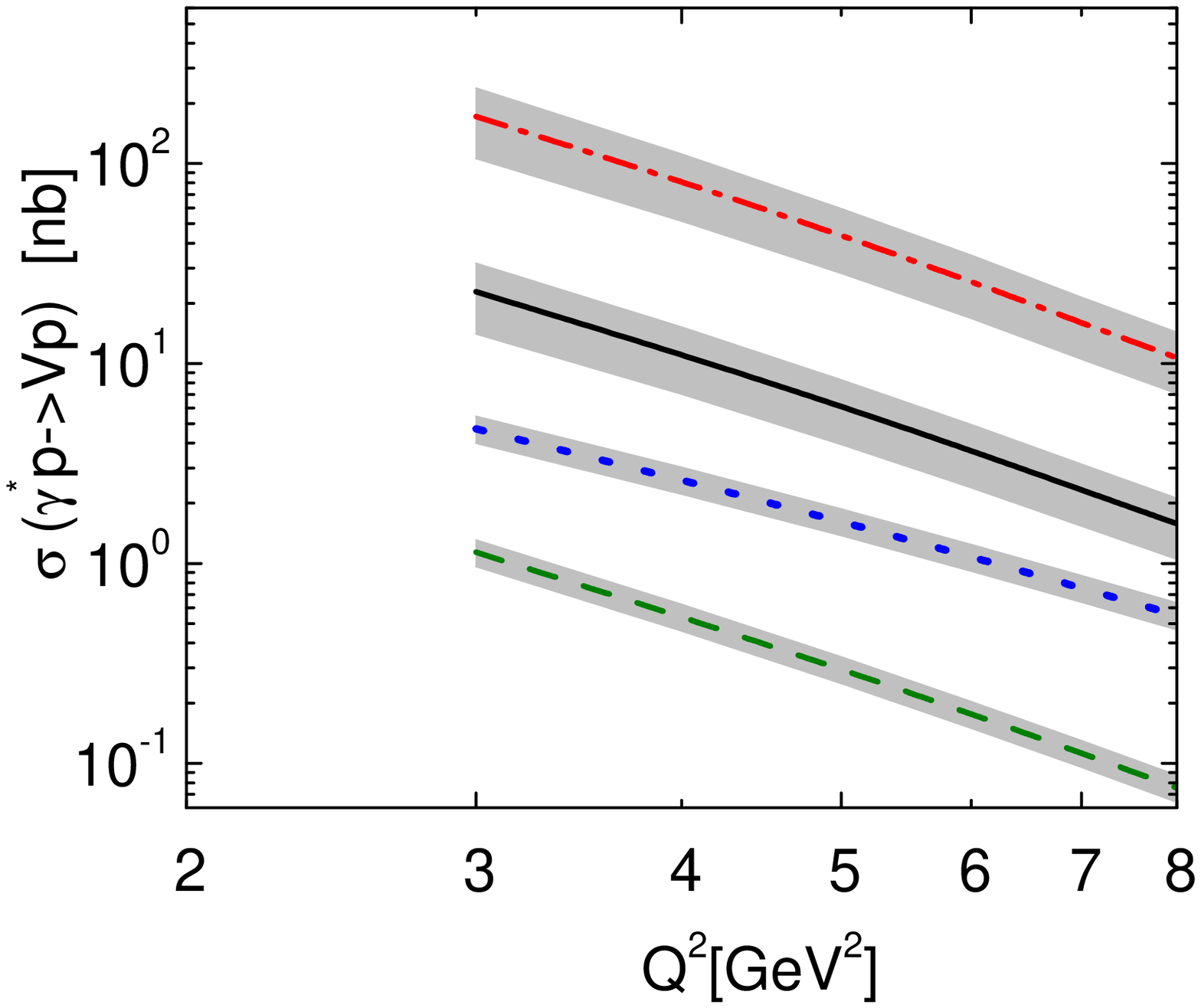}
\hspace*{0.05\tw}
\includegraphics[width=0.44\tw,bb= 37 345 508 743, clip=true]{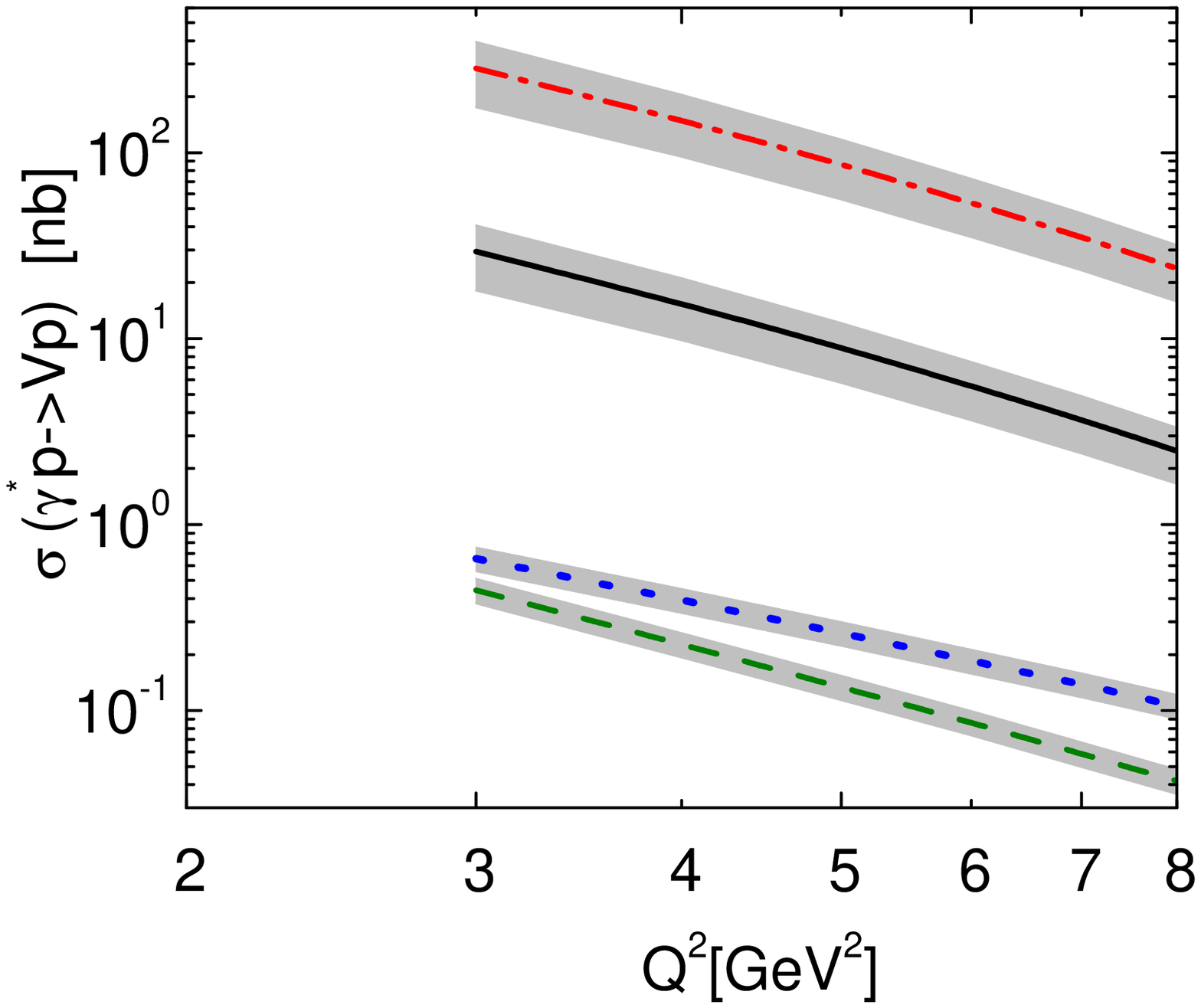}
\caption{The integrated cross section for vector-meson electroproduction  
  at $W=5$ (left) and $10\,\gev$ (right). The solid (dashed, dotted, 
  dash-dotted) line represents the results evaluated from  variant 1 
  for $\omega$ ($K^{*0}$, $\rho^+$, $\rho^0$) production. The shaded 
  bands indicate the theoretical uncertainties of our predictions. For 
  the $K^{*0}$ wave functions the Gegenbauer coefficients 
  $B_1^{K^{*0}L(T)}$ are taken to be zero.}
\label{fig:cross}
\end{figure}

Having specified various variants of the GPD $E$ and taking $H$ from
\ci{GK2} we are in the position to evaluate $A_{UT}$ for
electroproduction of various vector mesons. Of course we also obtain 
results for the cross sections (separated and unseparated). In
\ci{GK2,GK3} we have already compared the predictions for the cross
sections with experiment in great detail for the cases of $\rho^0$ and 
$\phi$ production and found generally excellent agreement with the
data from HERMES, HERA and FNAL in a large range of kinematics. 
Predictions for the unseparated integrated cross sections for $\omega$, 
$\rho^+$ and $K^{*0}$ electroproduction are shown in Fig.\
\ref{fig:cross} where, for comparison, also results for $\rho^0$
production are displayed. The theoretical uncertainties of our results
for the cross sections are estimated from the Hessian errors of the
CTEQ6 PDFs, cf.\ the discussion in \ci{GK2,GK3}. Due to neglected power
corrections of order $m^2/Q^2$ and $-t/Q^2$ and the possibility of
large higher order perturbative QCD corrections \ci{kugler07,ivanov08} we do not
provide results for $Q^2<3\,\gev^2$. 

The cross sections for $\rho^0$, $\phi$ and $\omega$ production
increase with energy at fixed $Q^2$ due to considerable contributions
from the gluonic subprocess $\gamma^* g\to Vg$ which grow $\propto
W^{4\delta(Q^2)}$ for $\xi\to 0$. This behaviour is to be contrasted
with the cross sections for $\gamma^* p\to K^{*0}\Sigma^+$ and 
$\gamma^*p\to\rho^+p$ to which the gluons do not contribute. These
cross sections are predicted to shrink with energy since the dominant 
valence quark contributions lead to $\sigma\propto W^{4(\alpha_{\rm val}(0)-1)}$ 
for $\xi\to 0$. The shrinkage is milder for $K^{*0}$ than for the
$\rho^+$ channel due to the strange quark contribution 
which has the same energy dependence as the gluon one.
Lack of data for these two processes prevent the verification of
our results as yet. It would be very interesting to learn whether 
the adopted $Q^2$-independent intercept of the valence quark Regge
trajectory is indeed required by experiment.

\begin{figure}[ht]
\includegraphics[width=0.46\tw,bb=37 327 533 754,clip=true]{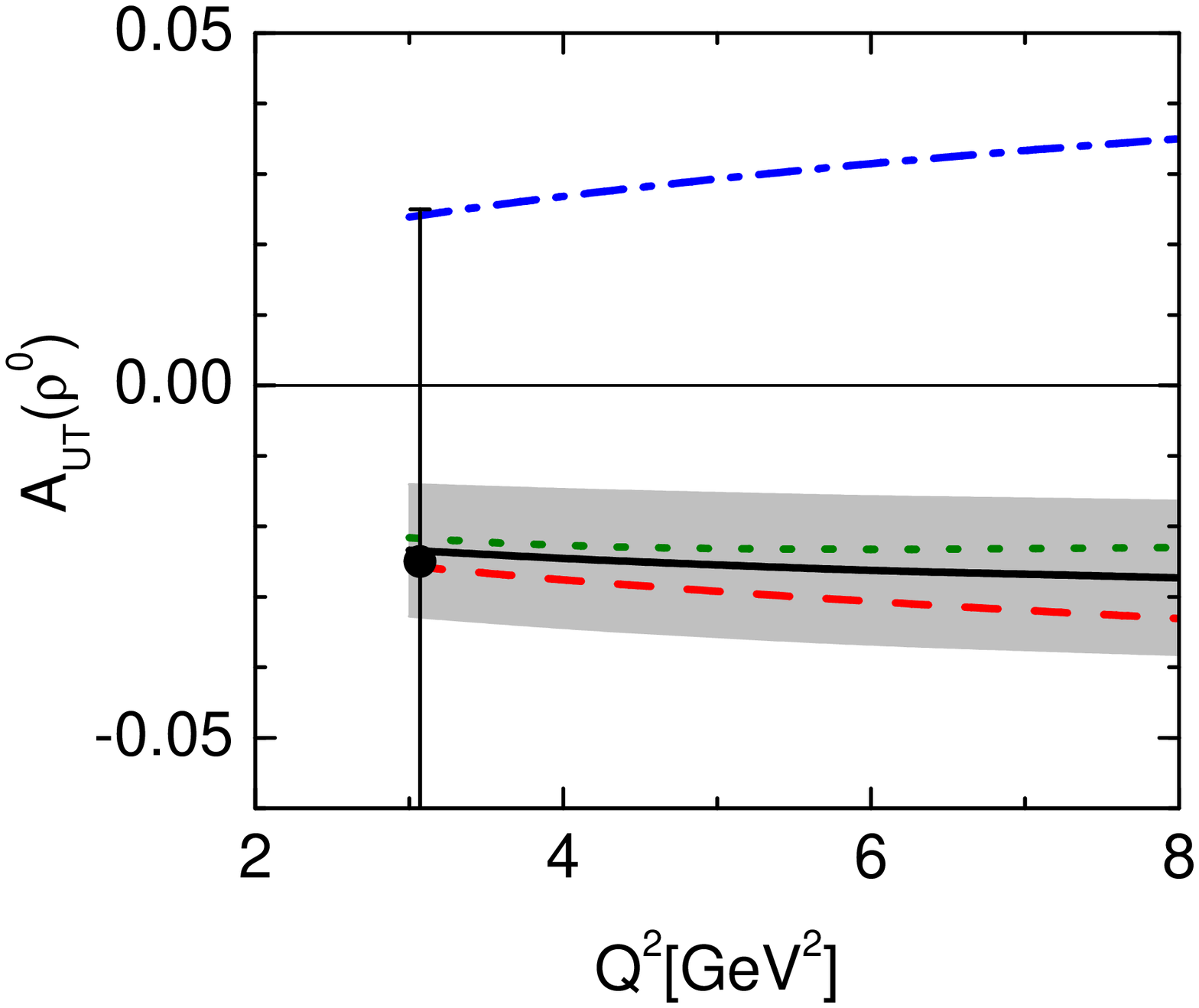}
\hspace*{0.05\tw}
\includegraphics[width=0.45\tw,bb= 42 327 530 744, clip=true]{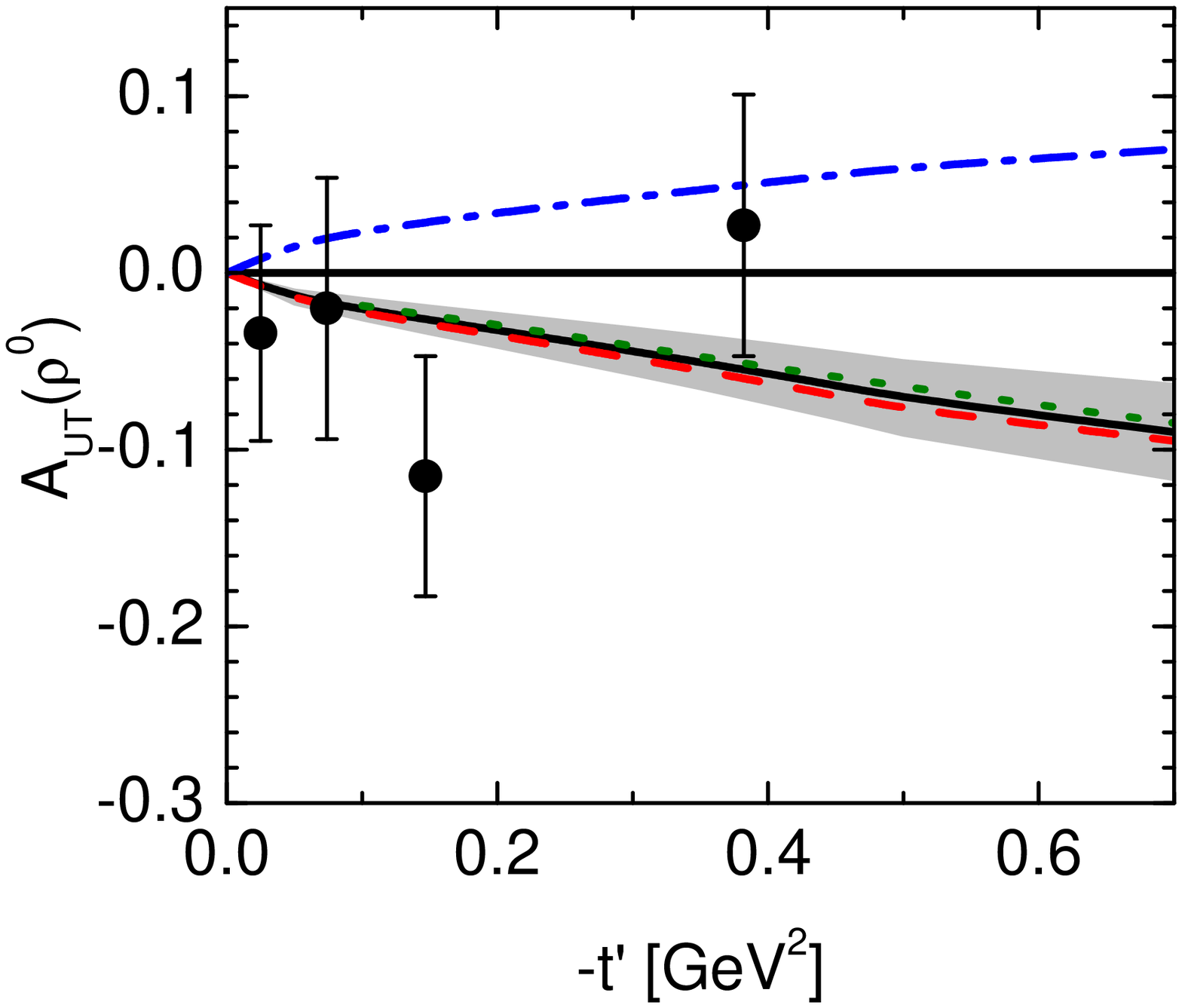}
\caption{The asymmetry $A_{UT}$ for $\rho^0$ production versus $Q^2$, 
  integrated upon $t$, (left) and versus $t'$ at $Q^2=3\,\gev^2$
  (right) at an energy of $5\,\gev$.  The solid (dashed, dotted,
  dash-dotted) line represents the results for variant 1 (2, 3,
  4). The shaded band indicates the theoretical uncertainty
  of the predictions for variant 1. The other variants have similar
  uncertainties. Preliminary data are taken from HERMES \ci{ami}, the
  $t$ dependent ones are at $Q^2=2\,\gev^2$.}
\label{fig:aut-rho}
\end{figure} 
In Fig.\ \ref{fig:aut-rho} we show the predictions for $A_{UT}$ for  
$\rho^0$ production at $W=5\,\gev$ integrated upon $0\leq -t \leq
0.5\,\gev^2$ ($t\simeq t'$ at small skewness) and versus $t$ at
$Q^2=3\,\gev^2$. The uncertainties of our 
results follow from those of the CTEQ6 PDFs and the parametric errors
of the GPD $E$. Note that the variants 2 and 3 are very similar to the 
default variant 1, i.e.\ the sea and gluon contributions cancel each
other to a large extent. The results obtained with variants 5 and 6
also lie within the error bands but are not shown here and in the following 
for better legibility of the figures. Only variant 4 exhibits a
distinctively different behaviour - $A_{UT}$ is now positive. Our
predictions obtained from the various variants agree with preliminary
data from HERMES \ci{ami}. Even variant 4 is not ruled out given the 
experimental and theoretical errors.
HERMES also provides $A_{UT}$ separately for longitudinally and 
transversally polarized  photons: $A_{UT}(\rho^0_L)=0.04\pm 0.12$ and 
$A_{UT}(\rho^0_T)=-0.08\pm 0.10$ at $W=5\,\gev$ and $Q^2=3.07\,\gev^2$.  
We find  $A_{UT}(\rho^0_L)\simeq A_{UT}(\rho^0_T)=-0.020$ for the
default variant. Although there is agreement within errors we stress
that the same sign is obtained for both these observables. The origin
of this fact can be traced back to the smallness of the phases
between the transverse and longitudinal amplitudes, about $3$
degrees, for both proton helicity flip and non-flip.
Consequences of this feature of our handbag approach for the spin
density matrix elements of the $\rho^0$ meson have been discussed 
in \ci{GK3}. In trend the predicted $t$ dependence is in agreement 
with the preliminary HERMES data \ci{ami} which are measured at the 
admittedly low value~\footnote{
We stress that in principle $A_{UT}$ can also be calculated at $Q^2$
lower than say $3\,\gev^2$. The trends of $A_{UT}$ to be seen from the
figures, simply propagate  down to smaller $Q^2$. What prevents us
from doing so is the fact that the theoretical uncertainties of the
predictions are not  under control at low $Q^2$ because of the
neglected power and higher order QCD corrections.}   
of $2\,\gev^2$ for $Q^2$.    

\begin{figure}[ht]
\includegraphics[width=0.45\tw,bb=31 325 534 755,clip=true]{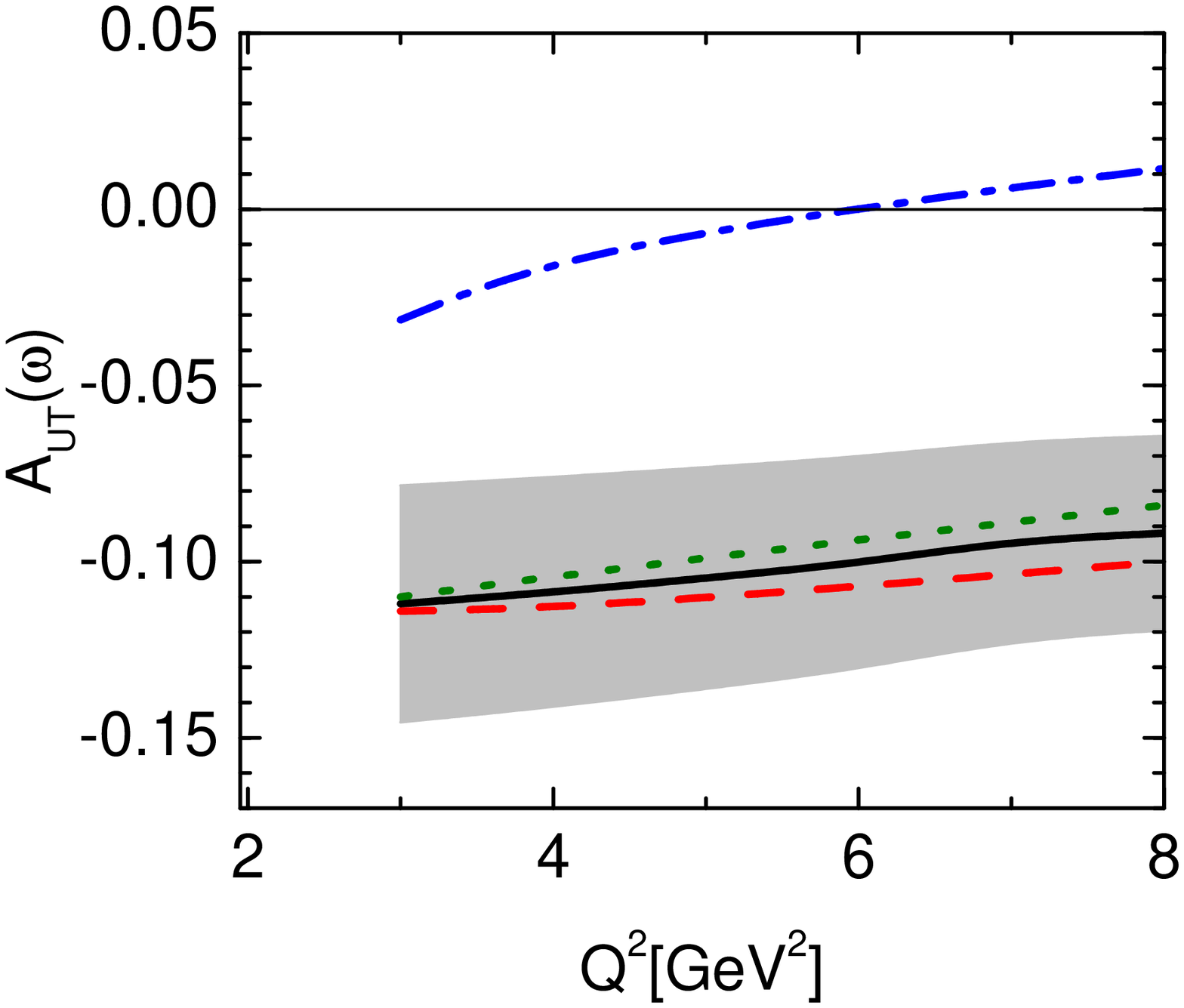}
\hspace*{0.1\tw}
\includegraphics[width=0.43\tw,bb=52 326 532 755, clip=true]{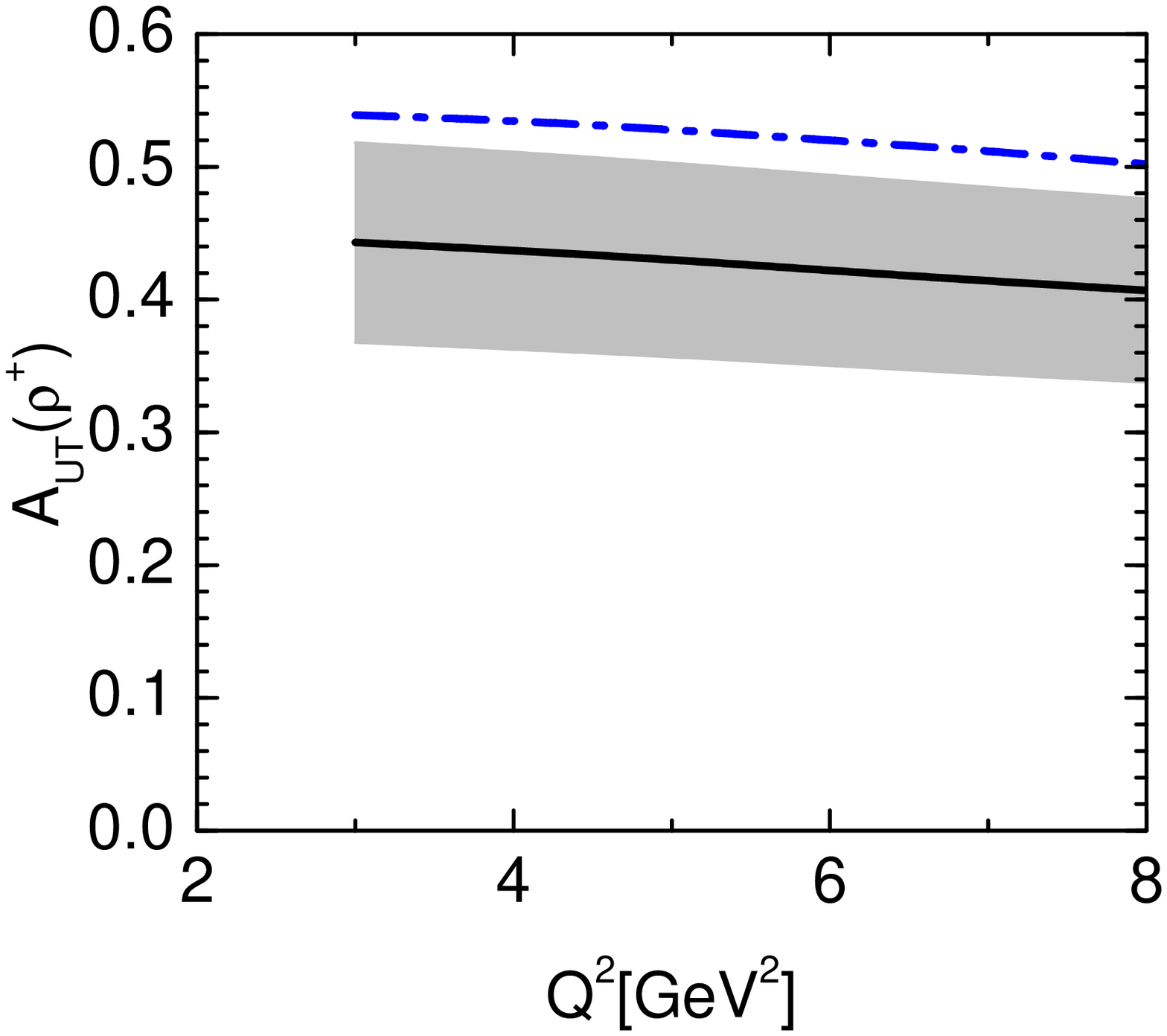}
\caption{As in Fig.\ \ref{fig:aut-rho} but for $\omega$ (left)
  and $\rho^+$ production (right).} 
\label{fig:aut-omega}
\end{figure}

The target asymmetry for $\omega$ and $\rho^+$ production at
the same energy is shown in Fig.\ \ref{fig:aut-omega}. For $\omega$ 
production $A_{UT}$ behaves similar to that for $\rho^0$ production 
except that it is about a factor of 5 larger in absolute value. 
This comes about from the flavor weigth factors \req{flavor} for the 
dominant valence quark contributions
\be
{\cal M}_{\mu -,\mu +}(\rho^0) \sim \langle e_u E^u_{\rm val} - e_d E^d_{\rm val}\rangle \qquad
{\cal M}_{\mu -,\mu +}(\omega) \sim \langle e_u E^u_{\rm val} + e_d E^d_{\rm val}\rangle \,. 
\ee
With $E^u_{\rm val}\simeq -E^d_{\rm val}$ as is indicated by the
anomalous magnetic moments of the quarks, $\kappa_a$, which fix the
normalization of $E^a_{\rm val}$, and corresponding effects in
the non-flip amplitudes the enhancement of $A_{UT}(\omega)$ over
$A_{UT}(\rho^0)$ in absolute value is evident.    

For $\rho^+$ production only $u$ and $d$ quark GPDs contribute. In
fact they occur in the combination
\be
F^u - F^d \simeq F^u_{\rm val} -  F^d_{\rm val}
\ee
where $F$ is either $H$ or $E$. The sea quark contributions cancel to
a large extent in the difference at small $\xi$ (cf.\ the discussion 
in \ci{GK2}). Therefore the variants 1, 2 and 3 fall together for this channel. 
Since $0<H^d_{\rm val} <H^u_{\rm val}$ and $E^u_{\rm
  val} \simeq -E^d_{\rm val}$ it is evident that the proton
helicity-flip amplitude is large in this case and leads to a very
large $A_{UT}(\rho^+)$, see Fig.\ \ref{fig:aut-omega}.

\begin{figure}[ht]
\begin{center}
\includegraphics[width=0.47\tw,bb=24 328 532 754,clip=true]{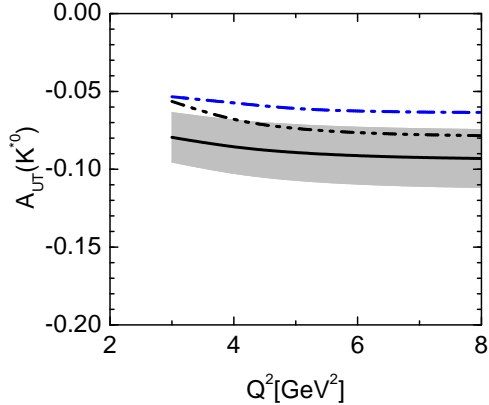}
\caption{The target asymmetry for $K^{*0}$ production at
  $W=5\,\gev$. The solid (dash-dotted) line represents
  our results for variant 1 (4) using $B_1^{K^{*0}L(T)}=0$. The
  dot-dot-dashed line are the results evaluated from variant 1 with
  $B_1^{K^{*0}L(T)}=0.1$.} 
\end{center}
\label{fig:aut-k0}
\end{figure}

Another interesting case is $K^{*0}$ production. Neither the  gluonic
GPDs contribute nor those for $u$ quarks; only the difference $F^d-F^s$
occurs. As a consequence the cross section is very small. On the other
hand, $A_{UT}$ is rather large in absolute value. This can be seen from
Fig.\ \ref{fig:aut-k0} where we also display results obtained from
variant 1 but using the value 0.1 for the first Gegenbauer
coefficient, $B_1^{K^{*0\, L(T)}}$, of the $K^{*0}$ wave function instead
of zero as employed for the other cases. The dependence on the Gegenbauer
coefficient is mild. Since the sea in $E$ is assumed to be flavor symmetric
the results for variants 1, 2 and 3 fall together.  

We do not display results for $\phi$ production. They are tiny,
typically less than $0.005$ in absolute value and for the default
variant even zero. The theoretical  uncertainties of our predictions
for $A_{UT}(\phi)$ are estimated to amount to about 0.01. The small
values of $A_{UT}(\phi)$ is not a surprise given the smallness of the
contributions from $E^g$ and $E^s$ and their partial cancellations for
variants 2 and 3. A preliminary HERMES result \ci{hermes-phi} of 
$A_{UT}(\phi)=-0.05\pm 0.12$ at $W\simeq 4.5\,\gev$ and $Q^2=1.9\,\gev^2$
is compatible with our results in trend although the experimental error is too
large for permitting any conclusion. The preliminary HERMES results 
\ci{hermes-phi} on $A_{UT}(\phi)$ for either longitudinally or transversally
polarized photons are also compatible with zero within large errors.

\begin{figure}[ht]
\includegraphics[width=0.46\tw,bb=42 326 529
746,clip=true]{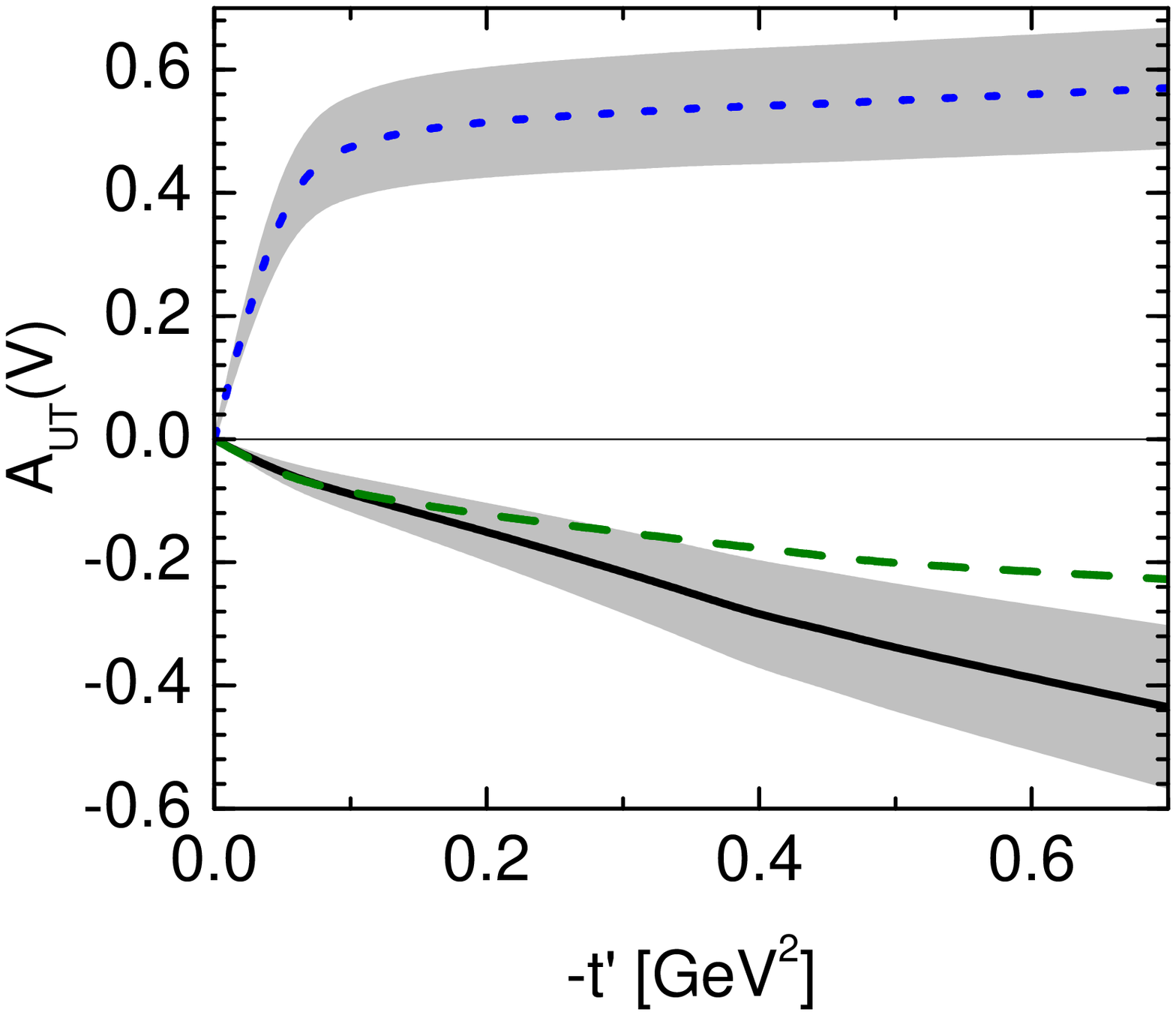}
\hspace*{0.05\tw}
\includegraphics[width=0.46\tw,bb=44 327 533 755,clip=true]{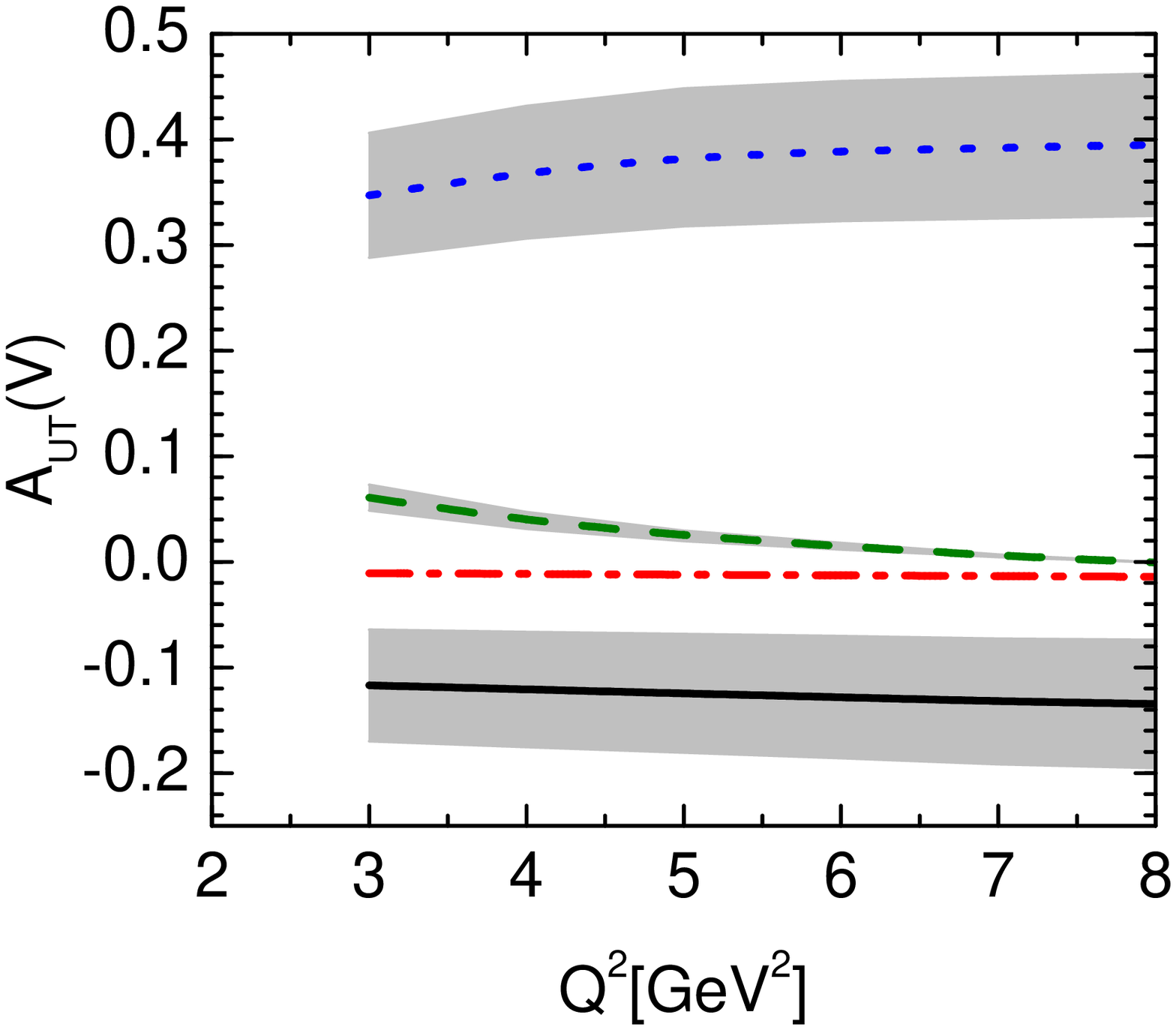}
\caption{The asymmetry $A_{UT}$ for various vector mesons versus $t'$
  at $W=5\,\gev$ and $Q^2=3\,\gev^2$ (left) and versus $Q^2$ at
  $W=10\,\gev$ integrated upon $t'$ (right), evaluated from
  variant 1. For notations refer to   Fig.\ \ref{fig:cross}.}  
\label{fig:aut-all}
\end{figure}

The $t$ dependence of $A_{UT}$ for the processes of interest,
evaluated from variant 1 at $W=5\,\gev$ and $Q^2=3\,\gev^2$ is shown  
in Fig.\ \ref{fig:aut-all}. While for $\omega$ and $K^{*0}$ production
it is similar to that of the $\rho^0$ case (see Fig.\
\ref{fig:aut-rho}) it is quite different for $\rho^+$ production. This
fact is a consequence of the large proton helicity flip amplitude
in the latter case. It provides a substantial contribution to the
cross section growing $\propto t'$, see Eq.\ \req{hel-amp}. Therefore, 
the $t$ dependence of $A_{UT}(\rho^+)$ is not approximately given by the
factor $\sqrt{-t'}/2m$ as for the other vector mesons. In Fig.\ 
\ref{fig:aut-all} we also show the $t$-integrated $A_{UT}$ at 
$W=10\,\gev$, again evaluated from variant 1. The observed energy 
dependence is understandable considering the Regge behaviour of the 
various GPDs at low $\xi$. 

Before closing this section a few comments are in order. Frequently it
is assumed that calculations to leading-twist accuracy of ratios
like $A_{UT}$ (see for instance \ci{kugler07,ossmann,GPV}) 
might provide realistic estimates despite possible failures
with the normalizations of the cross sections. This supposition relies 
on the assumption of a common $K$-factor for all amplitudes. In the 
modified perturbative approach which we are employing in the
calculation of the subprocess amplitudes, this is however not the case
with the exception of $\rho^0$ and $\phi$ production (see \ci{GK3}). The most
extreme example is $\omega$ production for which the leading-twist
result is about a factor of 2 larger in absolute value than our result
shown in Fig.\ \ref{fig:aut-omega}.

COMPASS is using a transversally polarized deuteron target for
electroproduction of flavor-neutral vector mesons. Suppose the
kinematics is chosen in such a way that the incoherent sum of proton
and neutron scattering is essentially measured~\footnote{
This may require large $Q^2$, see for instance \ci{EMC}.}.
In this situation and regarding that, at $W\simeq 10 \,\gev$, the
unpolarized cross sections for a proton and neutron target are about
equal (the gluonic contribution dominates, see \ci{GK2}) one has
\be
A_{UT}(V^0d) \simeq \frac12 \big[A_{UT}(V^0p) + A_{UT}(V^0n)\big]\,,
\ee
for $V^0=\rho^0, \omega, \phi$. Since the GPD $E$ is dominated by 
valence quarks with $E^u_{\rm val}\simeq - E^d_{\rm val}$ as we 
repeatedly mentioned, one can readily see that $A_{UT}(V^0d)$ is
approximately zero. This result is in agreement with preliminary
COMPASS data on $\rho^0$ production \ci{sandacz}. The treatment 
of coherent scattering is beyond the scope ot the present article.
  
\begin{figure}[ht]
\begin{center}
\includegraphics[width=0.45\tw,bb=29 328 533 755,clip=true]{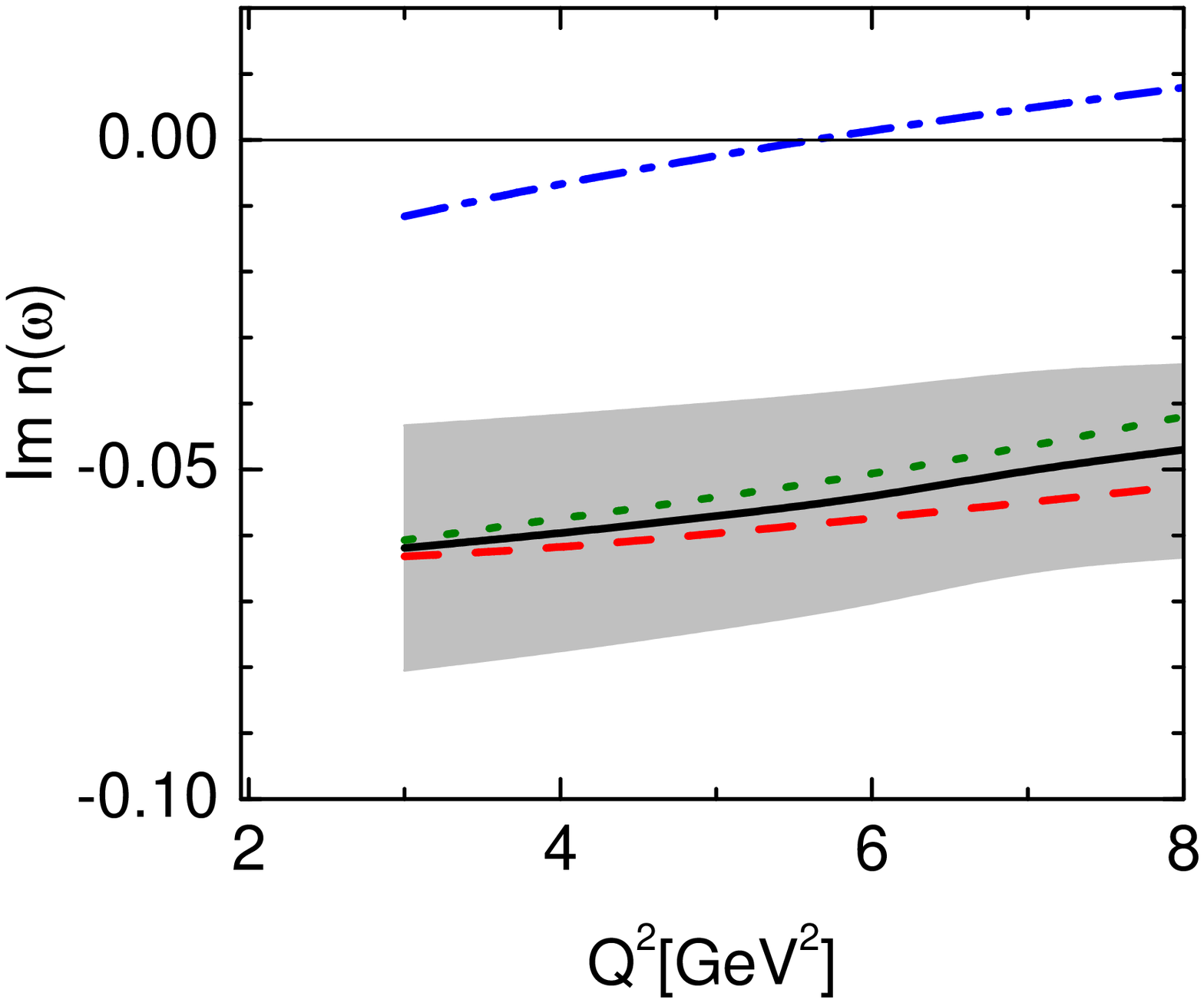}
\hspace*{0.05\tw}
\includegraphics[width=0.45\tw,bb=27 326 535 755,clip=true]{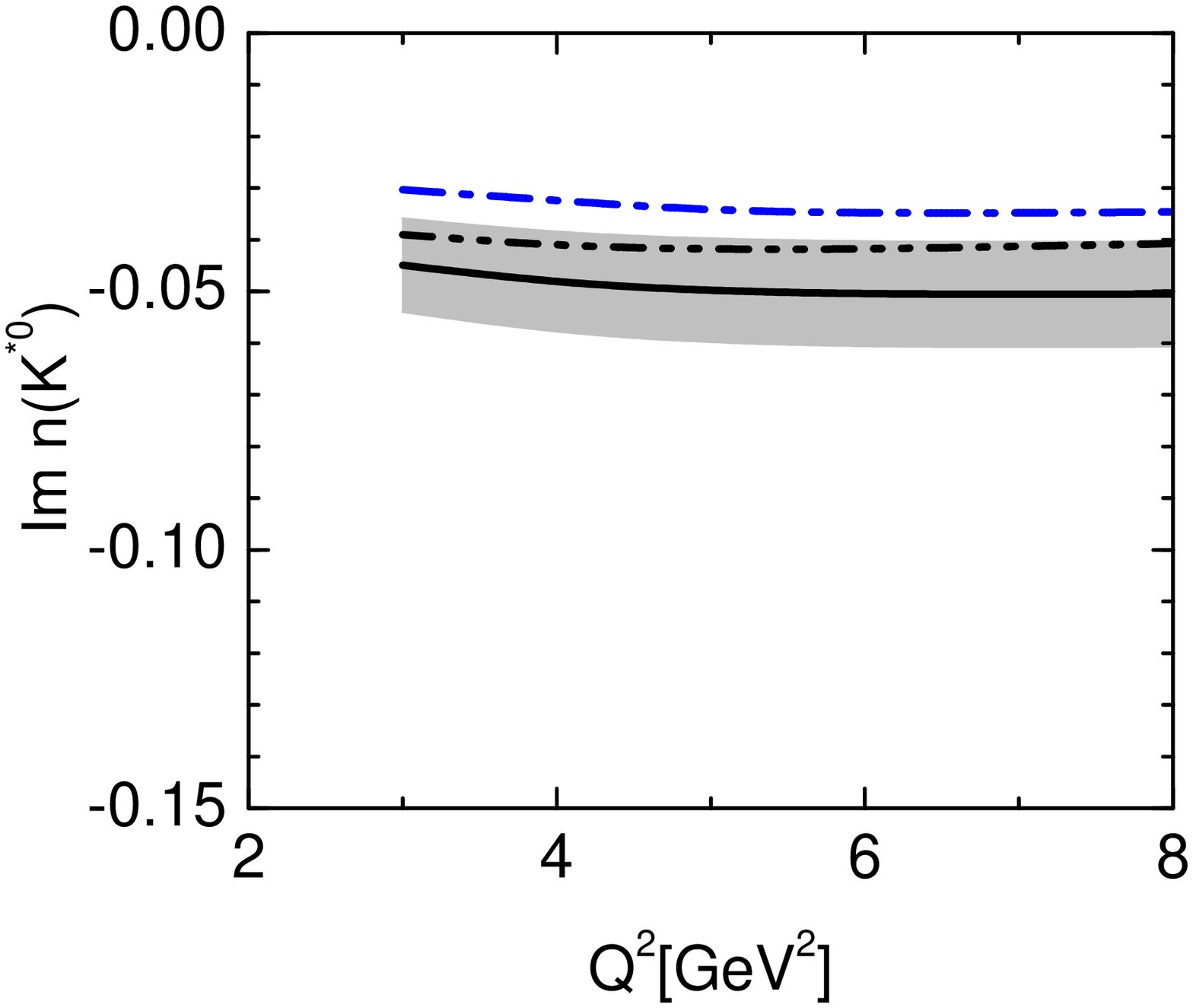}
\caption{left: Left: The SDME $Im\, n_{0+}^{0+}$, integrated upon $t$, for $\omega$
  (left) and $K^{*0}$ (right) production versus $Q^2$ at $W=5\,\gev$. For
  notations refer to Figs.\ \ref{fig:aut-rho} and \ref{fig:aut-k0}.}
\end{center}
\label{fig:sdme2}
\end{figure}

Recently the SDME formalism has been developed \ci{diehl07} for the
case of a proton target polarized perpendicular with respect to the
hadron plane. These SDMEs are denoted by $n^{\sigma\sigma'}_{\mu\mu'}$ 
and related to bilinear combinations of the amplitudes for helicities 
$\mu, \mu'$ and $\sigma, \sigma'$ of the virtual photon and the meson, 
respectively. Taking into account the amplitudes \req{hel-amp} only
the SDMEs $n^{00}_{00}$, $n^{++}_{++}$ and $n^{0+}_{0+}$ are non-zero. 
The first two are just $A_{UT}$ for longitudinally and transversally 
polarized virtual photons, respectively. Their sum $n^{++}_{++}+\veps
n^{00}_{00}$ is the unseparated target asymmetry that we discussed
above in great detail. These two SDMEs do therefore not provide any new
information in constrast to $n^{0+}_{0+}$ which is defined by
\be 
n^{0+}_{0+} \=- (n^{+0}_{+0})^*\=  \frac2{N_T+\veps N_L}\,  
    \Big[{\cal M}^N_{0-,0+}\,{\cal M}^{N*}_{++,++} - 
        {\cal M}^N_{0+,0+}\,{\cal M}^{N*}_{+-,++}   \Big]\,,
\label{eq:n}
\ee
and for which we show results for $\omega$ and $K^{*0}$ production in
Fig.\ \ref{fig:sdme2}. Results for $\rho^0$ production can be found in
\ci{GK3}.

\section{Concluding remarks}
In this study we have investigated the target asymmetry for
electroproduction of various vector mesons within the handbag approach. 
The hard subprocess amplitudes have been calculated employing the 
modified perturbative approach in which quark transverse degrees of 
freedom and Sudakov suppressions are taken into account. The GPDs are 
constructed from their forward limits combined with reggeized $t$
dependences with the help of double distributions. The GPD $H$ in 
particular has been modelled in our previous work \ci{GK1,GK2,GK3} and 
shown to provide very good fits of the data on the longitudinal and 
transverse cross sections for $\rho^0$ and $\phi$ production in a
large kinematical region. Here in this work we have taken $H$ which 
controls the proton helicity-non-flip amplitude for small skewness, as 
given and have concentrated ourselves on the construction of $E$. This 
GPD builds up the proton helicity-flip amplitude required for the 
calculation of $A_{UT}$. As $H$ it is constructed from double 
distributions and is constrained by the Pauli form factors of the
nucleon, positivity bound and the sum rule \req{E-sum-rule}. It turns 
out that for all variants we consider $E$ is dominated by the valence 
quarks. This feature of $E$ is to be contrasted with the behaviour of 
its counterpart $H$ for which the gluons play the most prominent role 
for not too large momentum fractions.

The present experimental and theoretical knowledge does not allow to
fix $E$ uniquely. Therefore, we have considered a number of variants from
which we have evaluated $A_{UT}$. It turns out that a particular
pattern is obtained for it. The sign and size of $A_{UT}$ is
characteristic of the flavor nature of the respective vector meson.
Hence, experimental data on $A_{UT}$ for various vector mesons even if
they are not very precise, may allow for a verification of the basic 
features of $E$ like the prominent role of the valence quarks, and
perhaps constrain its parameters tighter than it is actually the
case. The present still preliminary data from HERMES, measured at
rather small values of $Q^2$, are fully consistent with our predictions. 
Since on the other hand they suffer from large errors they do not rule 
out any of the variants we presented, only the extreme variant 4 is
slightly disfavored. Better data on $A_{UT}$ for $Q^2\gsim
3\,\gev^2$ are mandatory for a more accurate determination of
$E$. Such data may be provided by HERMES and COMPASS and perhaps from 
experiments performed at the upgraded JLab (see \ci{weiss08}). 

From the forward limits of $H$ and $E$ we have evaluated the angular
momenta of the partons by means of Ji's sum rule. We have found for
them a characteristic stable pattern which seems to be very hard to
change drastically within the approach we advocated for. The spin of
the proton is mainly built up by the $u$ quarks and the gluons with
only minor contributions from $d$ and $s$ ones. The connection between
$A_{UT}$ and the parton angular momenta seems to be quite general and
does not strongly depend on a particular model for the GPDs which
provide the link between these two quantities. Suppose the GPDs are 
smooth functions with no zeros except possibly at the end-points. The 
convolution of the GPD with the propagators which build up the 
amplitudes should then show the same trend in size as the second
moment of the GPD $E$, i.e.\ the contribution of $E$ to Ji's sum
rule. Moreover, with the exception of the $\rho^0$ case the $t$ 
dependence of $A_{UT}$ is dominated by the trivial factor 
$\sqrt{-t}/2m$ (see Eq.\ \req{hel-amp}) whose appearance is a 
consequence of angular momentum conservation. This implies that to a 
good approximation, $A_{UT}$ provides information on $E$ at $t\simeq
0$. In reality flavor decomposition of $E$ matters and renders the 
analysis more difficult. 

\section*{Acknowledgements} 
We thank  A. Borissov,  W.-D. Nowak, A. Sandacz and W. Vogelsang for 
discussions. We are also grateful to the HERMES collaboration
for permission to use preliminary data. This work has been supported
in part by the Heisenberg-Landau program and the Russian Foundation 
for Basic Research, grant 06-02-16215.


\vskip 10mm
\begin{thebibliography}{99}
\bibitem{GK1} S.~V.~Goloskokov and P.~Kroll,
  Eur.\ Phys.\ J.\ C {\bf 42}, 281 (2005)
  [hep-ph/0501242].

\bibitem{GK2} S.~V.~Goloskokov and P.~Kroll,
Eur.\ Phys.\ J.\  C {\bf 50}, 829 (2007)
  [hep-ph/0611290].

\bibitem{ami} A.\  Rostomyan and J.\ Dreschler [HERMES collaboration],
             arXiv:0707.2486[hep-ex].

\bibitem{ji:97} X.~D.~Ji,
Phys.\ Rev.\ Lett.\ {\bf 78}, 610 (1997) [hep-ph/9603249].

\bibitem{nowak} F.~Ellinghaus, W.-D.~Nowak, A.~V.~Vinnikov and Z.~Ye,
  Eur.\ Phys.\ J.\  C {\bf 46}, 729 (2006)
  [hep-ph/0506264].



\bibitem{GK3} S.~V.~Goloskokov and P.~Kroll,
Eur.\ Phys.\ J.\  C {\bf 53}, 367 (2008)
 [arXiv:0708.3569 [hep-ph]].

\bibitem{botts89} J.~Botts and G.~Sterman,
                   Nucl.~Phys.~B {\bf 325}, 62 (1989).

\bibitem{man} L.~Mankiewicz and G.~Piller,
                Phys.\ Rev.\ D {\bf 61}, 074013 (2000)
                [hep-ph/9905287];

\bibitem{teryaev} I.~V.~Anikin and O.~V.~Teryaev,
Phys.\ Lett.\ B {\bf 554}, 51 (2003)
[hep-ph/0211028].

\bibitem{Frankfurt:99}
  L.~L.~Frankfurt, P.~V.~Pobylitsa, M.~V.~Polyakov and M.~Strikman,
  Phys.\ Rev.\  D {\bf 60}, 014010 (1999)
  [hep-ph/9901429].

\bibitem{braun96} P.~Ball and V.~M.~Braun,
  Phys.\ Rev.\  D {\bf 54}, 2182 (1996)
  [hep-ph/9602323].

\bibitem{ball07}
  P.~Ball, V.~M.~Braun and A.~Lenz,
  JHEP {\bf 0708}, 090 (2007)
  [arXiv:0707.1201 [hep-ph]].

\bibitem{shifman}  M.~A.~Shifman and M.~I.~Vysotsky,
  Nucl.\ Phys.\  B {\bf 186}, 475 (1981).


\bibitem{diehl03} M.~Diehl,
  Phys.\ Rept.\  {\bf 388}, 41 (2003)
  [hep-ph/0307382].
 
\bibitem{sapeta} M.~Diehl and S.~Sapeta,
  Eur.\ Phys.\ J.\  C {\bf 41}, 515 (2005)
  [hep-ph/0503023].

\bibitem{kugler07} M.~Diehl and W.~Kugler,
  Eur.\ Phys.\ J.\  C {\bf 52}, 933 (2007)
  [arXiv:0708.1121 [hep-ph]].

\bibitem{DFJK4} M.~Diehl, T.~Feldmann, R.~Jakob and P.~Kroll,
  Eur.\ Phys.\ J.\  C {\bf 39}, 1 (2005)
  [hep-ph/0408173].

\bibitem{mus99} I.~V.~Musatov and A.~V.~Radyushkin,
Phys.\ Rev.\ D {\bf 61}, 074027 (2000) [hep-ph/9905376].

\bibitem{ossmann}J.~Ossmann, M.~V.~Polyakov, P.~Schweitzer, D.~Urbano and K.~Goeke,
  Phys.\ Rev.\  D {\bf 71}, 034011 (2005)
  [hep-ph/0411172].

\bibitem{ji:95}  X.~D.~Ji, J.~Tang and P.~Hoodbhoy,
  Phys.\ Rev.\ Lett.\  {\bf 76}, 740 (1996)
  [hep-ph/9510304].

\bibitem{pobyl} P.~Pobylitsa,
Phys.\ Rev.\ D {\bf  66}, 094002 (2002)
[hep-ph/0204337].

\bibitem{burkardt03} M.~Burkardt, 
Phys.\ Lett.\ B {\bf 582}, 151 (2004)
[hep-ph/0309116].

\bibitem{donnachie} A.\ Donnachie
  Phys.\ Lett.\  B {\bf 611}, 255 (2005)
  [hep-ph/0412085].

\bibitem{polyakov99} M.~V.~Polyakov and C.~Weiss,
  Phys.\ Rev.\  D {\bf 60}, 114017 (1999)
  [hep-ph/9902451].

\bibitem{hermes-xx} P.~Liebing  [HERMES Collaboration],
  AIP Conf.\ Proc.\  {\bf 915}, 331 (2007).
 
\bibitem{compass-xx} CM.~Alekseev {\it et al.}  [COMPASS Collaboration],
  arXiv:0802.3023 [hep-ex];
  E.~S.~Ageev {\it et al.}  [COMPASS Collaboration],
  Phys.\ Lett.\  B {\bf 633}, 25 (2006)
  [hep-ex/0511028].

\bibitem{cteq6} J.~Pumplin, D.~R.~Stump, J.~Huston, H.~L.~Lai, P.~Nadolsky and W.~K.~Tung,
  JHEP {\bf 0207}, 012 (2002)
  [hep-ph/0201195].

\bibitem{haegler} Ph.~H\"agler {\it et al.}  [LHPC Collaborations],
  arXiv:0705.4295 [hep-lat].

\bibitem{balitsky97} I.~Balitsky and X.~D.~Ji,
  Phys.\ Rev.\ Lett.\  {\bf 79}, 1225 (1997)
  [hep-ph/9702277].

\bibitem{barone98} V.~Barone, T.~Calarco and A.~Drago,
  Phys.\ Lett.\  B {\bf 431}, 405 (1998)
  [hep-ph/9801281].

\bibitem{BB} J.\ Bl\"umlein and H.\ B\"ottcher,
Nucl.\ Phys.\ B {\bf 636}, 225 (2002) [hep-ph/0203155]

\bibitem{florian} D.~de Florian, R.~Sassot, M.~Stratmann and W.~Vogelsang,
  arXiv:0804.0422 [hep-ph].

\bibitem{ivanov08} D.~Yu.~Ivanov, Proc.\ EDS07, Hamburg (2007).
 
\bibitem{hermes-phi} W.~Augustiniak [HERMES collaboration],
  talk presented at DIS08, London (2008).

\bibitem{GPV} K.~Goeke, M.~V.~Polyakov and M.~Vanderhaeghen,
  Prog.\ Part.\ Nucl.\ Phys.\  {\bf 47}, 401 (2001)
  [hep-ph/0106012].


\bibitem{EMC} J.~Ashman  {\it et al.} [EMC collaboration],
   Z.\ Phys.\ C {\bf 39}, 169 (1988).

\bibitem{sandacz} A. Sandacz, talk presented at the workshop on
'Electromagnetic Interactions with Nucelons and Nuclei', Milos (2007).


\bibitem{diehl07} M.~Diehl,
  JHEP {\bf 0709}, 064 (2007) [arXiv:0704.1565[hep-ph]].


\bibitem{weiss08} M.~Strikman and C.~Weiss,
  arXiv:0804.0456 [hep-ph].

\end{thebibliography}
\end{document}